# Skyrme-HFB deformed nuclear mass table

J. Dobaczewski[∗†], M.V. Stoitsov[†∗∗] and W. Nazarewicz[†∗]

[∗]*Institute of Theoretical Physics, Warsaw University ul. Hoża 69, PL-00681 Warsaw, Poland*
[†]*Department of Physics and Astronomy, The University of Tennessee Knoxville, TN 37996, USA,*
*Joint Institute for Heavy Ion Research, Oak Ridge, TN 37831, USA*
*Physics Division, Oak Ridge National Laboratory, P.O. Box 2008, Oak Ridge, TN 37831, USA*
[∗∗]*Institute of Nuclear Research and Nuclear Energy, Bulgarian Academy of Sciences*
*Sofia-1784, Bulgaria*

**Abstract.** The Skyrme-Hartree-Fock-Bogoliubov code HFBTHO using the axial (2D) Transformed Harmonic Oscillator basis is tested against the HFODD (3D Cartesian HO basis) and HFBRAD (radial coordinate) codes. Results of large-scale ground-state calculations are presented for the SLy4 and SkP interactions.

## 1. INTRODUCTION

The code HFBTHO [1] solves the self-consistent HFB equations by using the axial (2D) Transformed Harmonic Oscillator (THO) basis [2], which allows for a correct treatment of the single-quasiparticle wave function asymptotics. As discussed recently [3], the THO technique is a method of choice for performing massive nuclear structure calculations including weakly bound systems. In order to fully test the formalism, in the present study we present results obtained with the axial (2D) HFBTHO (v1.64) code compared to those obtained with two other codes: HFODD (v2.08i) [4], which uses a Cartesian (3D) Harmonic Oscillator (HO) basis and spherical (1D) HFBRAD [5], which uses a lattice of points in the radial coordinate.

In Ref. [3] we have published the first complete mass table of even-even nuclei obtained by using the THO method for the SLy4 Skyrme force [6]. Here we discuss one specific improvement of the method, and also present new results obtained with the SkP Skyrme force [7]. More details, including downloadable tables of ground-state properties can be found at `http://www.fuw.edu.pl/~dobaczew/thodri/thodri.html`.

## 2. TESTS

In this section, we discuss results of two numerical tests. First, by switching off the Local Scaling Transformation (LST) of THO, we run HFBTHO in the axial HO basis and test it against HFODD. For a given Skyrme interaction and zero-range, density-dependent pairing force, both codes should give exactly the same results. Since technical details of the inner structure of both codes are completely different, such calculations constitute an extremely stringent test of both codes.

Second, by switching the LST on, we could test the code HFBTHO against the spherical code HFBRAD [5]. Here, results of both codes cannot be *exactly* identical, because the phase spaces in which the solutions are obtained are significantly different.

Table 1 displays the results of test calculations performed for the SLy4 Skyrme interaction [6] and for the mixed zero-range pairing force [8]: $V(\vec{r})=V_0\left(1-\rho(\vec{r})/\rho_0\right)$ for $\rho_0$=0.32 fm$^{-3}$. The cutoff energy of $\varepsilon_{cut}$=60 MeV was used for summing up contributions of the HFB quasiparticle states to density matrices [2]. For a given phase space, the strength of the pairing force $V_0$ was adjusted so as to reproduce the experimental neutron pairing gap in $^{120}$Sn. The resulting values are $V_0$=−285.88, −284.10, and −284.36 MeV fm$^3$ for the HO (THO) bases of 680 and 3276 states, and for the radial box of $R_{box}$=30 fm, respectively. The radial HFBRAD calculations were performed with 300 points (i.e., the $\Delta r$=0.1 fm grid spacing), and the wave functions were included up to $j_{max}$=39/2. We checked that even with $j_{max}$=33/2, all energies were stable within 1 eV. The nucleon-mass and elementary-charge parameters were fixed at $\hbar^2/2m$=20.73553 MeV fm$^2$ and $e^2$=1.439978 MeV fm, respectively.

Table 1 displays the following quantities: $N_0$ is the maximum number of the HO oscillator quanta included in the basis (for the deformed basis we give the numbers of quanta in the perpendicular ($N_\perp$) and axial ($N_z$) directions); $N_{st}$ is the number of the lowest deformed HO states included in the basis; $N_n^{qp}$ and $N_p^{qp}$ are the numbers of (doubly degenerate) neutron and proton quasiparticle states with equivalent single-particle energies [2] below the cutoff energy $\varepsilon_{cut}$; $b_\perp$ and $b_z$ are the oscillator constants in the perpendicular and axial directions; $\lambda_n$ and $\lambda_p$ are the neutron and proton Fermi energies, which, for vanishing pairing correlations, are taken as the s.p. energies of the last occupied states; $\Delta_n$ and $\Delta_p$ are the average pairing gaps [7]; $R_n$ and $R_p$ are the rms radii; $Q_n$ and $Q_p$ are the quadrupole moments $\langle 2z^2-x^2-y^2\rangle$; $\varepsilon_n^{gs}$ and $\varepsilon_p^{gs}$ are the s.p. energies of the most bound neutron and proton states; $\Sigma_n^\varepsilon$ and $\Sigma_p^\varepsilon$ are sums of the canonical energies weighted by the corresponding occupation probabilities; $E_n^{pair}$ and $E_p^{pair}$ are the pairing energies; $E_n^{kin}$ and $E_p^{kin}$ are the kinetic energies; $E_{cen}$ and $E_{SO}$ are the energies corresponding to the central and spin-orbit parts of the Skyrme energy density functional; $E_{dir}$ and $E_{exc}$ are the direct and exchange parts of the Coulomb energy; and $E_{stab}$ is the stability energy characterizing the level of self-consistency. In the code HFODD, $E_{stab}$ is estimated from the sum of s.p. energies [9]; in the code HFBTHO $E_{stab}$ is estimated from the maximum difference of all matrix elements of s.p. potentials calculated in two consecutive iterations; and in the code HFBRAD it is calculated as a variance of the total binding energy, $E_{tot}$, over the last five iterations.

Calculations for $^{208}$Pb yield a spherical solution with vanishing pairing gaps. HFBTHO and HFODD give the total binding energies that differ by 627 eV, and this difference can be (primarily) traced back to the direct Coulomb energy. We have checked that without the Coulomb interaction, this difference decreases to 202 eV. The axial-basis HFBTHO calculation gives a very small total quadrupole moment of 39 $\mu$b. This suggests that the THO basis generates a slight deviation from the spherical symmetry due to a different numerical treatment of $z$- and $\perp$-direction. In this respect, HFODD calculations should be considered more accurate.

Calculations for $^{168}$Er performed within a spherical HO basis, $b_\perp$=$b_z$, yield a well-deformed and weakly paired prolate ground state. Here, the total binding energies and quadrupole moments obtained within HFBTHO and HFODD differ only by 32 eV and 5 $\mu$b, respectively. When the same calculation is performed in a deformed HO basis, $b_\perp \neq b_z$, the differences grow to 5.1 keV and 207 $\mu$b, respectively. Again, without the Coulomb interaction, the difference in the total binding energy is only 96 eV. It is seen that by employing the deformed basis, the binding energy decreases, as expected.

**TABLE 1.** (Color online) Benchmark results of the HFB calculations performed for the SLy4 interaction and mixed $\delta$ pairing. All energies are in MeV, lengths in fm, and quadrupole moments in barns. Boldface colored digits differ between the HFBTHO and HFODD/HFBRAD calculations. See text for details.

| Nucleus: | $^{208}$Pb | | $^{168}$Er | | $^{168}$Er | | $^{120}$Sn | |
|---|---|---|---|---|---|---|---|---|
| Code: Basis: | HFBTHO 2D-HO | HFODD 3D-HO | HFBTHO 2D-HO | HFODD 3D-HO | HFBTHO 2D-HO | HFODD 3D-HO | HFBTHO 2D-THO | HFBRAD Radial |
| $N_0$ | 14 | 14 | 14 | 14 | $N_\perp$=13, $N_z$=17 | $N_\perp$=13, $N_z$=17 | 25 | n.a. |
| $N_{st}$ | 680 | 680 | 680 | 680 | 680 | 680 | 3276 | n.a. |
| $N_n^{qp}$ | 532 | 532 | 489 | 489 | 497 | 497 | 924 | 4260 |
| $N_p^{qp}$ | 481 | 481 | 448 | 448 | 451 | 451 | 855 | 4003 |
| $b_\perp$ | 2.2348121 | 2.2348121 | 2.1566616 | 2.1566616 | 2.0581218 | 2.0581218 | 2.0390141 | n.a. |
| $b_z$ | 2.2348121 | 2.2348121 | 2.1566616 | 2.1566616 | 2.3681210 | 2.3681210 | 2.0390141 | n.a. |
| $\lambda_n$ | −8.114095 | −8.114020 | −6.936061 | −6.936058 | −6.943872 | −6.943858 | −8.016795 | −8.018081 |
| $\lambda_p$ | −8.810501 | −8.810445 | −7.156485 | −7.156477 | −7.152114 | −7.152007 | −11.107284 | −11.107777 |
| $\Delta_n$ | 0 | 0 | 0.394570 | 0.394578 | 0.392326 | 0.392327 | 1.244750 | 1.244648 |
| $\Delta_p$ | 0 | 0 | 0.390601 | 0.390605 | 0.397728 | 0.397746 | 0 | 0 |
| $R_n$ | 5.619758 | 5.619757 | 5.357578 | 5.357578 | 5.360037 | 5.360044 | 4.730466 | 4.730184 |
| $R_p$ | 5.460080 | 5.460090 | 5.225538 | 5.225539 | 5.227218 | 5.227231 | 4.593884 | 4.593653 |
| $Q_n$ | −0.000022 | 6.6E-11 | 11.473921 | 11.473920 | 11.567875 | 11.567983 | −0.001055 | 0 |
| $Q_p$ | −0.000017 | 4.7E-11 | 7.880228 | 7.880224 | 7.930128 | 7.930227 | −0.000631 | 0 |
| $\varepsilon_n^{gs}$ | −58.001139 | −58.001145 | −56.014966 | −56.014973 | −55.996356 | −55.996370 | −55.756516 | −55.755837 |
| $\varepsilon_p^{gs}$ | −44.042810 | −44.042814 | −44.422148 | −44.422167 | −44.486154 | −44.486271 | −46.629670 | −46.631739 |
| $\Sigma_n^{\varepsilon}$ | −3009.265452 | −3009.264720 | −2401.023343 | −2401.023305 | −2401.701865 | −2401.698888 | −1667.965633 | −1668.063705 |
| $\Sigma_p^{\varepsilon}$ | −1678.791400 | −1678.790238 | −1439.480739 | −1439.480826 | −1439.922261 | −1439.913577 | −1123.812244 | −1123.857483 |
| $E_n^{pair}$ | 0 | 0 | −1.716956 | −1.717024 | −1.703028 | −1.703045 | −12.467146 | −12.466964 |
| $E_p^{pair}$ | 0 | 0 | −1.528611 | −1.528643 | −1.584308 | −1.584480 | 0 | 0 |
| $E_n^{kin}$ | 2525.991268 | 2525.991925 | 1974.613878 | 1974.613824 | 1973.986024 | 1973.980663 | 1340.457995 | 1340.668648 |
| $E_p^{kin}$ | 1334.854760 | 1334.854465 | 1118.313614 | 1118.313442 | 1118.495643 | 1118.487818 | 830.735396 | 830.848077 |
| $E_{cen}$ | −6194.978513 | −6194.978930 | −4944.027994 | −4944.027545 | −4943.869108 | −4943.856093 | −3475.705844 | −3476.043789 |
| $E_{SO}$ | −96.374920 | −96.375003 | −80.186775 | −80.186826 | −80.216433 | −80.214900 | −49.167364 | −49.196956 |
| $E_{dir}$ | 827.607126 | 827.607885 | 602.810399 | 602.810352 | 602.694020 | 602.697867 | 366.472441 | 366.503834 |
| $E_{exc}$ | −31.248467 | −31.248462 | −25.935909 | −25.935905 | −25.935633 | −25.935528 | −19.102496 | −19.103705 |
| $E_{stab}$ | 8.1E-09 | 3.5E-11 | 1.0E-08 | 3.4E-06 | 9.6E-09 | 3.8E-06 | 9.9E-09 | 8.8E-08 |
| $E_{tot}$ | −1634.148747 | −1634.148120 | −1357.658354 | −1357.658322 | −1358.132823 | −1358.127702 | −1018.777019 | −1018.790854 |

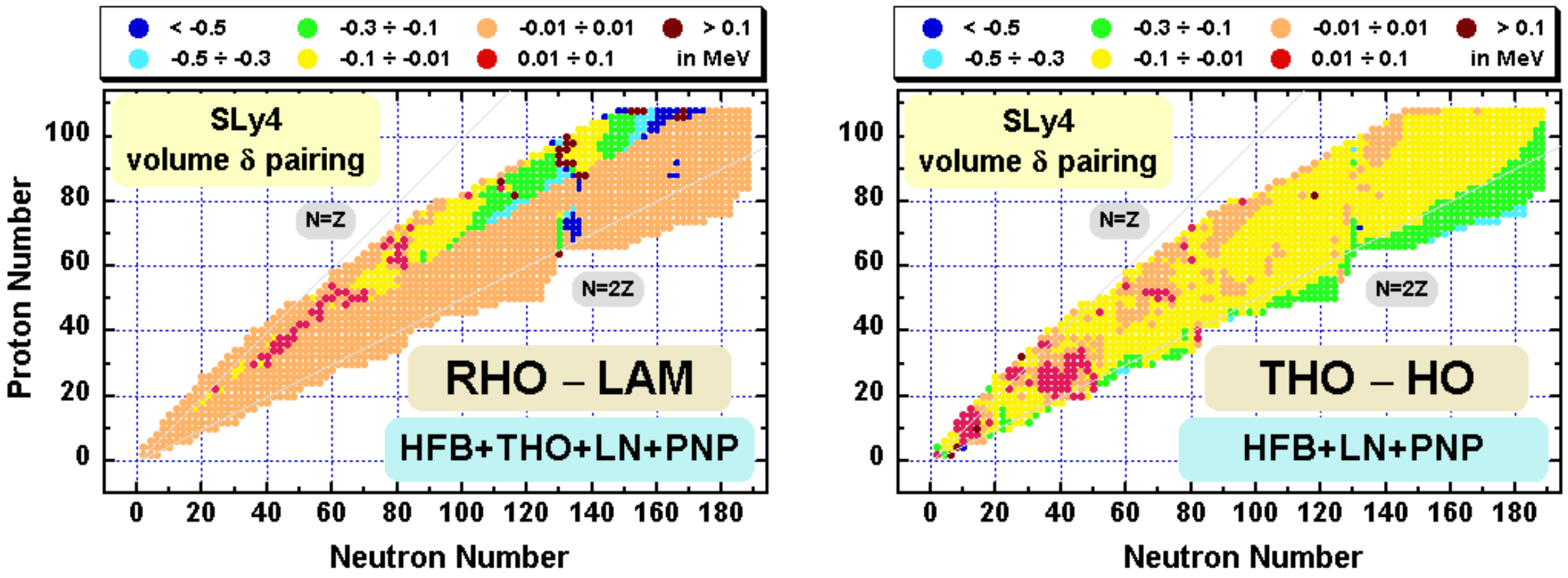

**FIGURE 1.** (Color online) Left: differences in $E_{tot}$ obtained in HFBTHO by using LST based on RHO or LAM conditions. Right: differences between $E_{tot}$ obtained in THO and HO bases. Calculations were performed using the SLy4 interaction with volume pairing and 20 oscillator shells. Lipkin-Nogami method followed by the exact particle-number projection was used to correct for the particle number nonconservation in HFB.

Comparison with the coordinate-space code HFBRAD for $^{120}$Sn shows that $E_{tot}$ in HFBTHO is correct up to 14 keV for $N_0$=25. However, the kinetic energy still differs by as much as 221 keV, which is compensated by a similar difference in the interaction energy. Within the HO basis and $N_0$=25, the corresponding differences are larger: 41 and 337 keV. The analogous differences obtained for $N_0$=20 are 142 and 1103 keV (THO), and 152 and 964 keV (HO), respectively. Nevertheless, the above comparison shows that the $N_0$=20 calculations yield $E_{tot}$ with a precision of a couple of hundred keV.

## 3. MASS TABLES

The LST employed in Ref. [3] was based on HO densities corrected in the asymptotic region by the contribution from the lowest-energy quasiparticle. Since a common LST has to be carried out for both neutrons and protons, for each nucleus one is forced to make a decision whether the LST is to be based on neutron or proton density. In Ref. [3] we used a prescription (referred to as LAM) that the neutron densities were used for $\lambda_n \geq \lambda_p$ and *vice versa*. In this work, we use the condition $\rho_n(R_{min}) \geq \rho_p(R_{min})$, where $R_{min}$ is the point where the neutron or proton logarithmic density has a minimum as a function of $r$. In practice, the above condition, dubbed RHO, does not depend on whether the neutron or proton $R_{min}$ is considered.

In Fig. 1 (left panel) we show the differences in $E_{tot}$ obtained in HFBTHO by using the LST condition employing the Fermi energies (LAM) [3] or the densities (RHO). One can see that in the majority of neutron-rich nuclei both prescriptions lead to identical results. However, in many proton-rich nuclei the new prescription decreases binding up to about 500 keV, and for some medium-mass proton-rich nuclei the RHO method *decreases* binding by up to 100 keV. This latter effect is due to a better description of asymptotics in the pairing channel, which leads to extended pairing fields and reduced pairing energies [10]. The right panel of Fig. 1 shows differences in $E_{tot}$ obtained in THO and HO bases. In most nuclei, by using the THO basis, one obtains a small energy gain of up to 10 keV. This grows to $\sim$500 keV for the very neutron-rich systems. Again, in lighter nuclei, a better asymptotics may lead to a reduced binding. In fact, our results show that improvements in density profiles at large distances cannot be treated variationally. First, $E_{tot}$ is quite insensitive to the precise description of nucleonic densities in outer nuclear regions. Second, due to the pairing-space cutoff, the pairing energy is not reacting variationally on the improvement of the wave function.

Figures 2 and 3 present HFBTHO results obtained with the SLy4 and SkP Skyrme forces. It is obvious that without further improvements these traditional Skyrme forces describe nuclear masses rather poorly. The rms deviations between calculated and measured masses are as large as 3.14 MeV for SkP and 5.10 MeV for SLy4, respectively, as compared to about 0.70 MeV deviations obtained for forces fitted specifically to masses (see Ref. [11] for a review). Moreover, pronounced kinks obtained at magic numbers suggest that the quality of the description of (semi)magic and open-shell systems is not the same. This may point to a need to systematically include dynamical zero-point corrections [12]. Work in this direction is in progress.

## ACKNOWLEDGMENTS

This work was supported in part by the Polish Committee for Scientific Research (KBN); by the Foundation for Polish Science (FNP); by the U.S. Department of Energy under Contract Nos. DE-FG02-96ER40963 (University of Tennessee), DE-AC05-00OR22725 with UT-Battelle, LLC (Oak Ridge National Laboratory), and DE-FG05-87ER40361 (Joint Institute for Heavy Ion Research); and by the National Nuclear Security Administration under the Stewardship Science Academic Alliances program through DOE Research Grant DE-FG03-03NA00083.

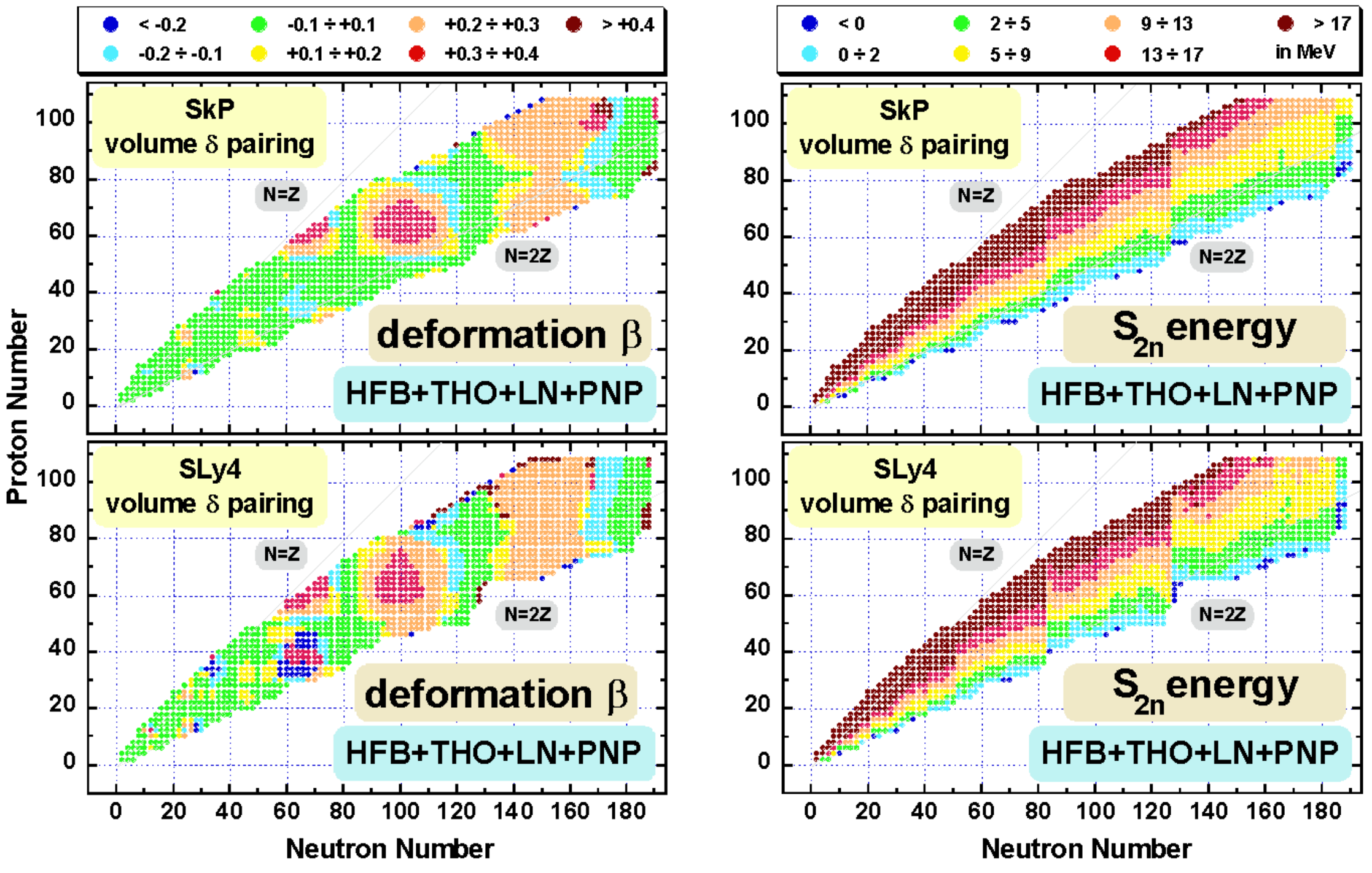

**FIGURE 2.** (Color online) Ground-state deformations $\beta$ (left) and two-neutron separation energies $S_{2n}$ (right) obtained within HFBTHO using SkP (top) and SLy4 (bottom) interactions.

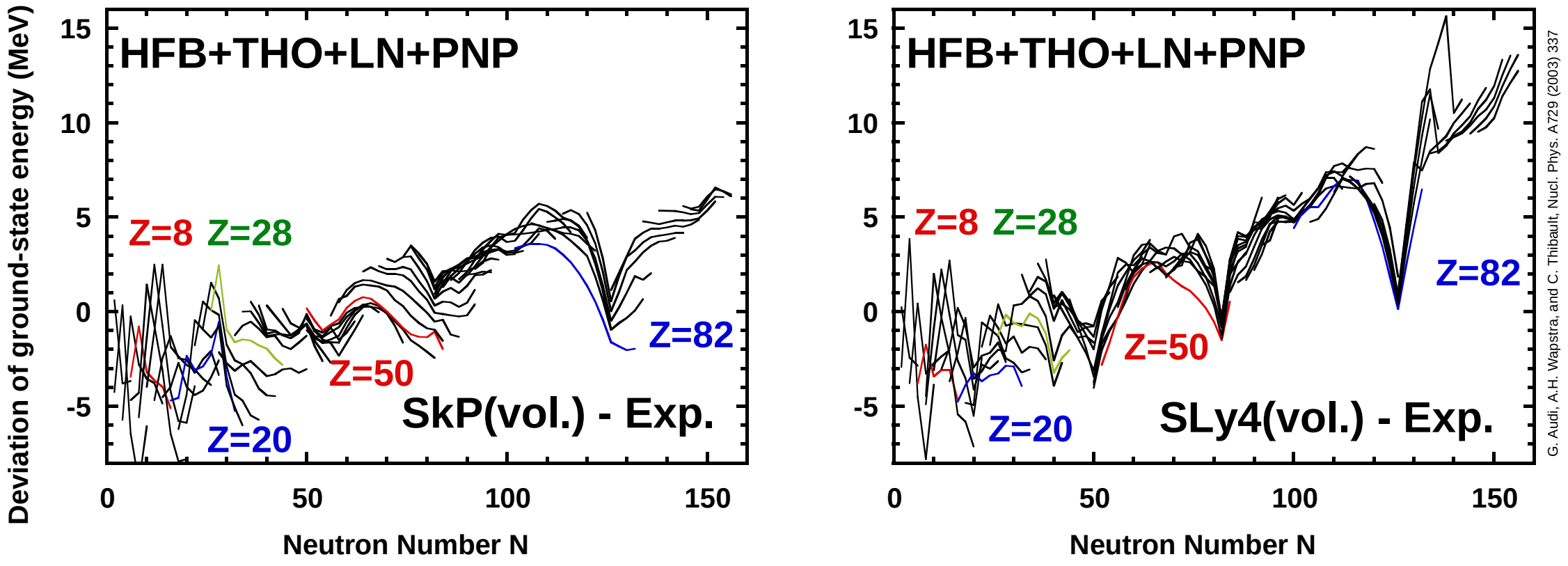

**FIGURE 3.** (Color online) Deviations of ground-state HFBTHO energies from experiment [13] for SkP (left) and SLy4 (right) interactions. Positive values correspond to underbound nuclei. No corrections beyond mean field were included.